\documentclass[11pt]{article}
\usepackage[preprint]{acl}

\usepackage{times}
\usepackage{latexsym}
\usepackage[T1]{fontenc}
\usepackage[utf8]{inputenc}
\usepackage{microtype}
\usepackage{inconsolata}
\usepackage{graphicx}
\usepackage{booktabs}
\usepackage{amsmath}
\usepackage{amsfonts}
\usepackage{nicefrac}
\usepackage{multirow}
\usepackage{subcaption}
\usepackage{placeins}
\usepackage{xcolor}

\title{P2Skill: Privacy Preserving Skill Distillation for Cloud-Local LLM Inference Systems}

\author{
  Myunghoon Ryu, Geunpyo Park, Sungjoon Lee, XinYu Piao, Jong-Kook Kim\\
  Department of Electrical and Computer Engineering\\
  Korea University\\
  Seoul, Republic of Korea\\
  \texttt{\{ryumh10,gppark,ssungjoon,xypiao97,jongkook\}@korea.ac.kr}
}
\date{}

\begin{document}
\maketitle

\begin{abstract}
Cloud-local LLM inference can use the reasoning capability of large cloud models, but cloud-bound requests must exclude personally identifiable information (PII).
Existing privacy-preserving methods modify the prompt or the model, which may distort contextual semantics or require additional training.
This paper proposes P2Skill, a prompt-based skill distillation method in which a local small language model (SLM) autonomously performs four skills.
A cloud LLM iteratively refines the skills from execution failures to let the local SLM generalize beyond memorized PII patterns, and the final skills are selected at the lowest validation leakage within set bounds.
P2Skill requires no privacy-specific fine-tuning or learned auxiliary detectors.
On real user queries, P2Skill on the best of three local SLMs transmits fewer annotated identifiers than every existing privacy-preserving method that sends text to the cloud and preserves quality more often than the closest of them in leakage, although fully local answering preserves quality more often.

\end{abstract}

\section{Introduction}
\label{secintro}

Recent large language models (LLMs) have found extensive deployment in personal agents, workplace automation, and decision support systems, embedding natural language inference into an expanding range of privacy-sensitive workflows~\citep{wang2024surveyagents}. Cloud-based inference has emerged as the predominant deployment paradigm for sustaining the performance demands of such applications, attracting considerable research and industrial attention~\citep{kwon2023vllm, hurst2024gpt4o}.
A fundamental concern inherent to this paradigm lies in the necessity of transmitting user input prompts to remote servers, which frequently contain personally identifiable information (PII)~\citep{mireshghallah2024trust, siyan2025papillon, das2025security}.
The consequent risk of PII exposure is particularly severe in domains requiring strict data confidentiality, such as healthcare~\citep{singhal2023med, hhs2024hipaa}, legal services~\citep{cui2023chatlaw}, and financial advising~\citep{wu2023bloomberggpt, ftc2024glba}.
Minimizing this leakage while maintaining inference quality constitutes a central problem in the deployment of cloud-based LLM services.

\begin{figure*}[t]
\centering
\includegraphics[width=\textwidth]{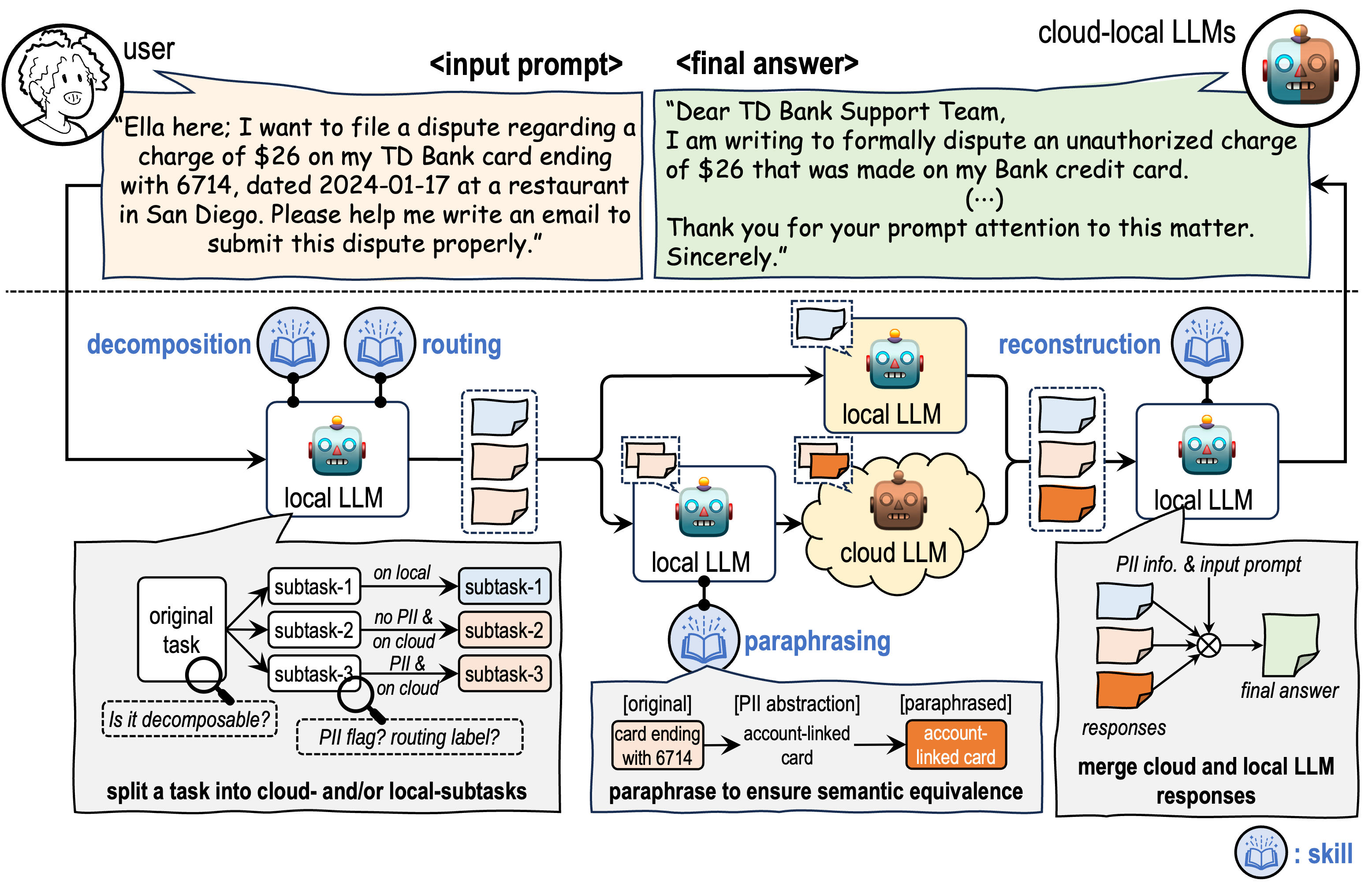}
\caption{Overview of the P2Skill pipeline.}
\label{fig:pipeline}
\end{figure*}

Prior work has explored several directions to protect PII contained in user prompts before transmission to cloud servers.
Cryptographic inference~\citep{giladbachrach2016cryptonets, mohassel2017secureml, mishra2020delphi} guarantees confidentiality but requires specialized primitives, model retraining for encrypted operations, and complex multi-party infrastructure.
Named-entity-recognition-based de-identification~\citep{presidio, zaratiana2024gliner, nvidia2025glinerpii, deusser2025resource} applies learned detectors that strip identifiable spans, but redacted prompts lose task-critical context~\citep{zeng2025privacyrestore, yang2025robust}.
Local differential privacy (LDP) methods~\citep{mai2024split, zhan2025prismprivacyawareroutingadaptive} perturb tokens before transmission, but the injected noise degrades linguistic coherence and the privacy-utility trade-off remains an inherent limitation~\citep{duchi2013local, acharya2020context}.
Cloud-local LLM inference systems process easy tokens or requests on the device and forward harder ones to the cloud LLM~\citep{jin2024ce, hao2024hybrid, li2025collaborative}, and privacy-aware systems keep sensitive content on the device by rewriting the prompt~\citep{siyan2025papillon} or by routing each prompt by its privacy risk~\citep{zhan2025prismprivacyawareroutingadaptive}.
However, the privacy module of these privacy-aware systems is a separate trained or prompt-optimized component and can require per-deployment adaptation.
Reconstructing the original semantic content from the privacy-redacted prompt also requires an additional trained restoration module~\citep{zeng2025privacyrestore} or an auxiliary mechanism deployed on the local device.

To minimize privacy leakage and maintain inference quality in cloud-local LLM inference systems, this paper introduces P2Skill, which distills privacy-preserving skills and applies them to a local small language model (SLM).
A skill is a reusable natural-language prompt that encapsulates a specific capability of an LLM~\citep{khot2023decomposed, li2026skillsbench, anthropic2026agentskills, openai2026skills}.
The two design choices that distinguish P2Skill from learned anonymizers and perturbation-based methods are (i) prompt-only operation on a frozen SLM that is designed to transmit only PII-free content to the cloud and (ii) closed-loop skill refinement whose memorization check rejects sample-specific rewrites.
P2Skill is tested on three SLMs and two benchmarks, and on real user queries, P2Skill on its best local SLM transmits fewer annotated identifiers than every compared baseline that sends text to the cloud and preserves quality more often than the closest of them in leakage, although fully local answering preserves quality more often.

\section{Related Work}
\label{sec:related}

Language models can emit memorized training-data identifiers~\citep{carlini2021extracting, lukas2023analyzing, kim2023propile}, but these studies do not concern cloud-bound prompts.
Split-and-Denoise adds LDP noise to representations~\citep{mai2024split}, and PrivacyRestore removes privacy spans and restores their information by server-trained vectors~\citep{zeng2025privacyrestore}, but the noise can degrade inference quality and the restoration can distort task-relevant semantics.
PAPILLON~\citep{siyan2025papillon} is a privacy-conscious delegation pipeline whose prompts are optimized~\citep{khattab2024dspy}, and PRISM~\citep{zhan2025prismprivacyawareroutingadaptive} combines entity-level sensitivity estimation, soft routing, and adaptive LDP.
Both rely on privacy-specific optimized, learned, or hand-set components that can require re-optimization, retraining, or recalibration for a new domain or PII distribution.
Unlike these methods, P2Skill decomposes each prompt and routes its sub-tasks by prompt-based skills of a frozen local SLM, detects PII by a skill and a deterministic identifier matcher instead of a learned detector or an LDP budget, and needs skill refinement on annotated prompts but no retraining or recalibration for a new domain.

\begin{figure*}[t]
\centering
\includegraphics[width=\textwidth]{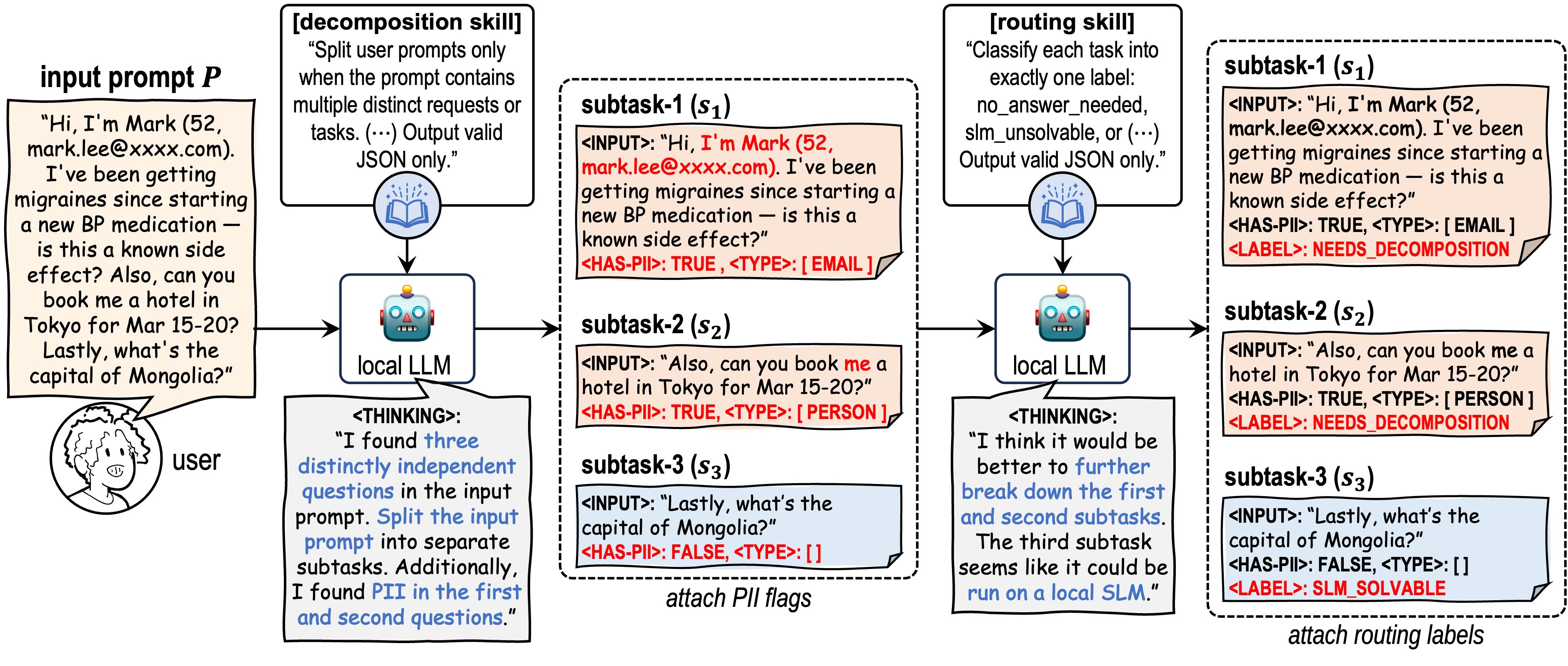}
\caption{Privacy-preserving decomposition (skill \#1) and PII-aware routing (skill \#2).}
\label{fig:stage1-2}
\end{figure*}

\section{Proposed Method}
\label{sec:method}

\subsection{Overview}

In P2Skill (Figure~\ref{fig:pipeline}), the user prompt is processed on the local SLM by four prompt-based skills.
The first skill decomposes the prompt into sub-tasks and flags each sub-task that contains PII.
The second skill attaches a routing label to each sub-task, and a deterministic dispatcher uses the label and the PII flag to route the sub-task to the local SLM or the cloud LLM.
Before transmission, the third skill paraphrases any cloud-bound sub-task that still contains detected PII.
The fourth skill reconstructs the sub-task results into a single final response.
Cloud-bound requests pass a deterministic identifier matcher that rejects known PII patterns.
The cloud LLM is designed to receive only PII-free requests.
Revisions of the skills come from a cloud-LLM-guided refinement loop, a stage-wise tuning procedure, and second initial versions (Section~\ref{sec:distill}).

\subsection{Pipeline Stages}
\label{sec:stages}

\paragraph{Privacy-preserving decomposition (skill \#1).}
The local SLM first uses the decomposition skill to split the user prompt $P$ into $k$ sub-tasks $\{s_1,\ldots,s_k\}$, where each sub-task captures one self-contained unit of work and carries a flag indicating whether it contains PII (Figure~\ref{fig:stage1-2}).
If the prompt contains independent questions or requests, the local SLM splits the prompt at those boundaries.
A prompt that consists of a single request remains intact.
The PII categories that the decomposition skill detects cover names, contact and location details, account and ID numbers, and personal facts that identify a user context, and their exact list differs between the skills of the three SLMs.
Public figures, organizations, and general factual content are not flagged unless the surrounding context reveals private information.
The decomposition skill is tuned to prefer a false flag to a missed identifier.
A false flag leads the sub-task to paraphrasing, to the local SLM, or to skipping at a possible cost to inference quality, whereas a missed identifier that the identifier matcher does not recognize propagates straight to the cloud and breaks the privacy protection.

\paragraph{PII-aware routing (skill \#2).}
The local SLM uses the routing skill to label each sub-task as solvable by the SLM, not solvable by the SLM, needing decomposition, or needing no answer.
A PII-flagged sub-task that the routing skill labels as needing no answer, such as context or metadata of the request, is skipped.
Let $\hat{s}_i$ denote the paraphrased rewrite of $s_i$ produced by skill \#3, and let $\textsc{Match}(\cdot)$ denote the deterministic identifier matcher, which flags any text that matches one of its regular expressions for e-mail and IP addresses, ZIP codes, phone, card, account, ID, and passport numbers, titled names, Korean names followed by an honorific, and Korean addresses.
The PII flag is $p_i{=}1$ if skill \#1 flags $s_i$ or gives no valid flag for it or if $\textsc{Match}(s_i){\neq}\text{clean}$, and an output of skill \#1 that cannot be parsed yields one sub-task that holds the whole prompt, has $p_i{=}1$, and is never skipped.
The routing rule $\textsc{Route}(s_i)$ is
\begin{equation}
\label{eq:route}
\resizebox{0.86\columnwidth}{!}{$\left\{\begin{array}{@{}l@{\ }l@{}}
\text{cloud (direct)} & p_i{=}0, \\
\text{cloud (paraphrased)} & p_i{=}1,\ \textsc{Match}(\hat{s}_i){=}\text{clean}, \\
\text{local SLM} & \text{otherwise}.
\end{array}\right.$}
\end{equation}
As an exception to Eq.~(\ref{eq:route}), a sub-task that has $p_i{=}1$ is answered on the local SLM without paraphrasing if the routing skill labels it as solvable by the SLM or if its identifier density exceeds a fixed threshold (at least four pattern matches in fewer than 400 characters or at least three pattern types in fewer than 250 characters), above which paraphrasing tends either to leave residual identifiers in the rewritten text or to remove the context required for an accurate answer.
Every other sub-task, including one labeled as not solvable by the SLM or as needing decomposition, follows Eq.~(\ref{eq:route}).

\begin{figure}[t]
    \centering
    \includegraphics[width=\columnwidth]{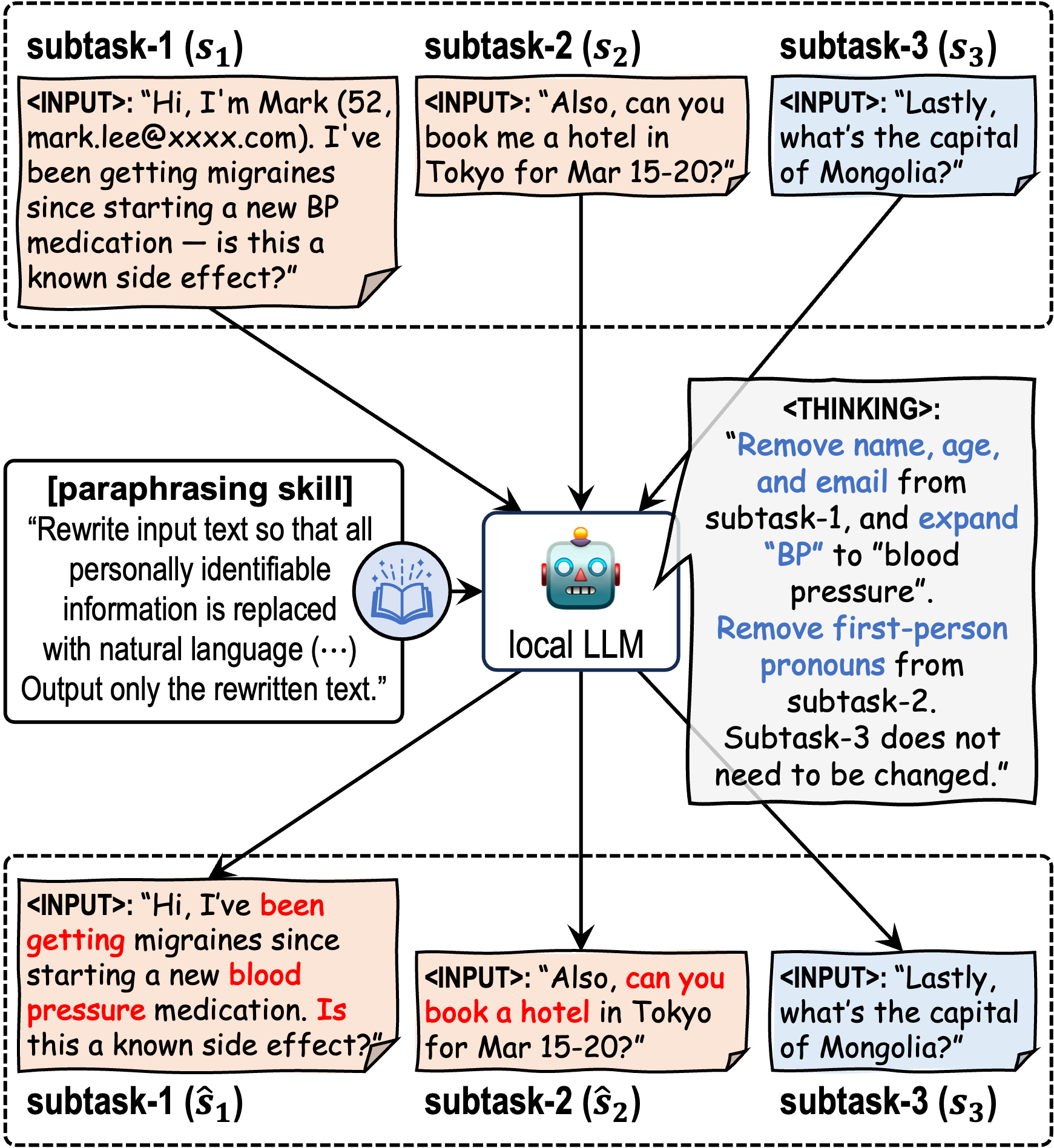}
    \caption{PII paraphrasing (skill \#3).}
    \label{fig:stage3}
\end{figure}

\paragraph{PII paraphrasing (skill \#3).}
The local SLM uses the paraphrasing skill to rewrite any cloud-bound sub-task $s_i$ that still contains detected PII (Figure~\ref{fig:stage3}).
The paraphrase $\hat{s}_i$ reaches the cloud only if it satisfies the substring-exclusion constraint
\begin{equation}
    \hat{s}_i {=} \textsc{Paraphrase}(s_i),\ p \not\sqsubseteq \hat{s}_i\ \forall p \in \textsc{PII}(s_i),
\end{equation}
where each $p \in \textsc{PII}(s_i)$ is the literal string of a raw identifier that the identifier matcher detects in $s_i$, and $p \sqsubseteq \hat{s}_i$ denotes that $p$ appears as a substring of $\hat{s}_i$.
The rewrite is designed to preserve role, intent, and document structure and to remove every detected identifier.
Removing identifiers necessarily changes the surface form of the request, and a careless rewrite can drop task-critical context and lower the quality of the cloud answer.
The skill itself cannot guarantee equivalence to the original because some information is intentionally discarded.
The refinement loop in Section~\ref{sec:distill} aims to close this gap by rewriting the paraphrasing skill to make future rewrites better retain task-critical context while removing every detected identifier.

\begin{figure}[t]
    \centering
    \includegraphics[width=\columnwidth]{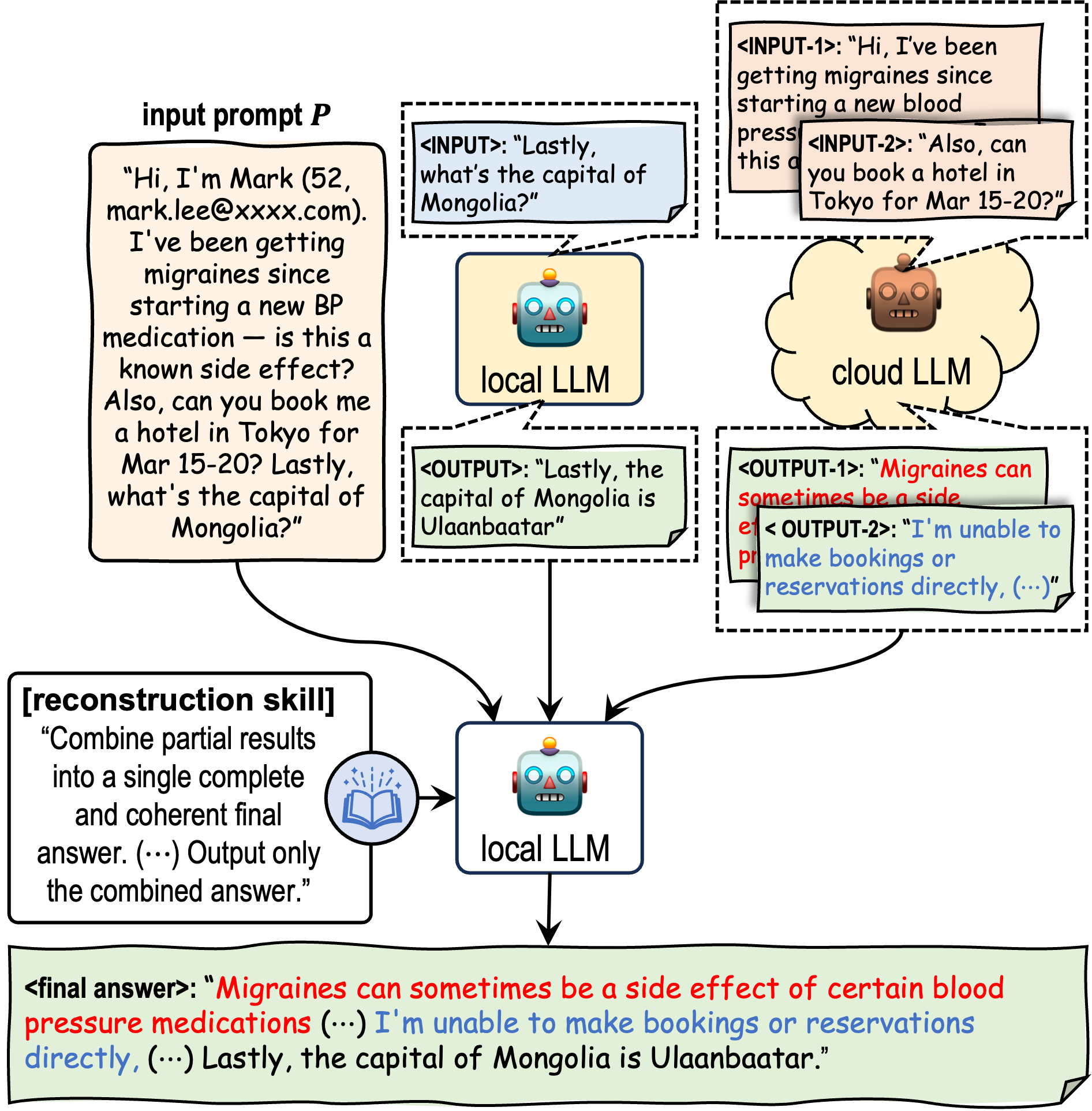}
    \caption{Final response reconstruction (skill \#4).}
    \label{fig:stage4}
\end{figure}

\begin{figure*}[t]
\centering
\includegraphics[width=\textwidth]{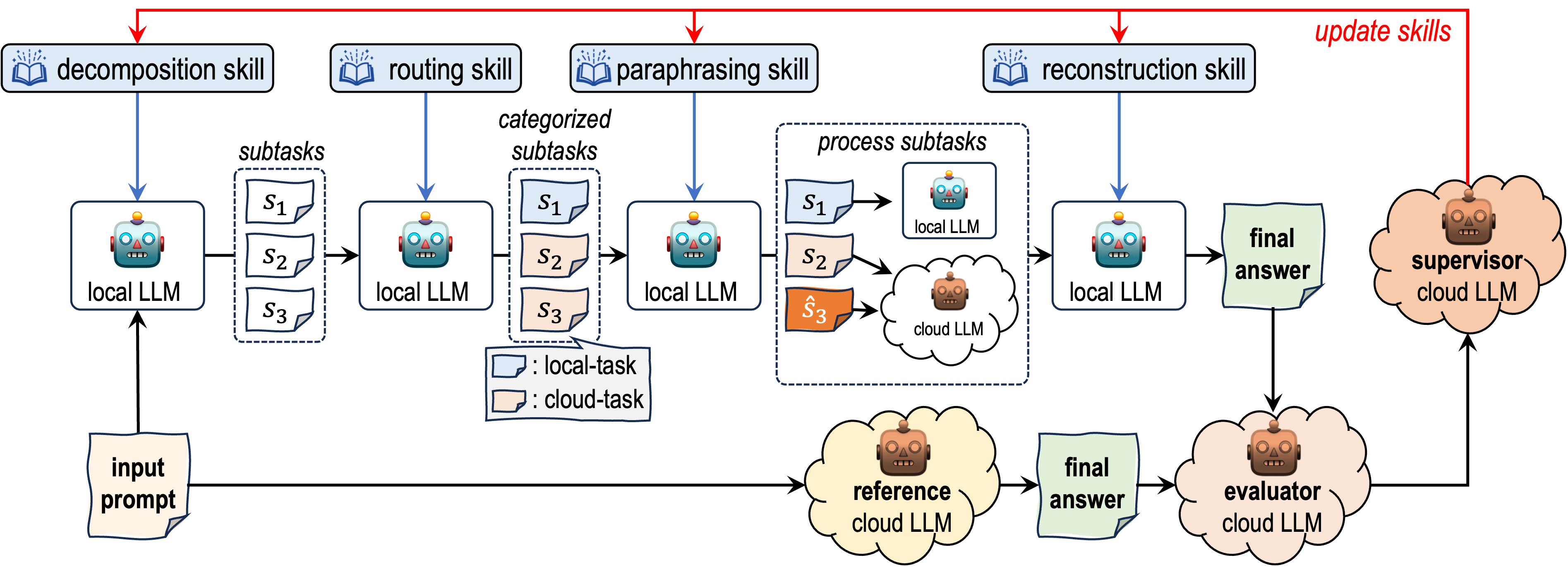}
\caption{Iterative refinement loop for skill distillation.}
\label{fig:distill}
\end{figure*}

\paragraph{Final response reconstruction (skill \#4).}
At the end of the pipeline, the local SLM uses the reconstruction skill to synthesize the final response from the original user prompt $P$ and the set of sub-task results $\{r_1,\ldots,r_k\}$, where $r_i$ denotes the answer produced for sub-task $s_i$ (Figure~\ref{fig:stage4}).
If a prompt yields a single answer and no paraphrased sub-task of the prompt reaches the cloud, this answer is the final response without reconstruction.
If no sub-task of a prompt yields an answer, for example because every sub-task is skipped, the final response is empty.
Otherwise, the local SLM composes an answer that integrates the available sub-task results, preserves concrete facts, and follows the format that the prompt asks for.
Wherever a sub-task result uses a generic description, the local SLM puts back the specific detail from $P$, and the user's own identifiers may appear in the final output even though the skills aim to exclude them from any cloud-bound request.

\subsection{How to Distill the Skills}
\label{sec:distill}

Figure~\ref{fig:distill} shows the skill distillation procedure, in which a cloud LLM guides the refinement as reference, evaluator, and supervisor, encoding the resulting knowledge into prompt-based skills rather than into the local SLM's weights.
An AI assistant writes the initial skill prompts from human instructions on the stage output format, behavioral constraints, and privacy policy.
The refinement loop revises the decomposition, paraphrasing, and reconstruction skills in rounds of $T$ iterations and uses a quality threshold $\tau$ to decide whether a scored response counts as a success or a failure.

\paragraph{Iterative refinement loop.}
Each iteration samples a mini-batch $B_t$ from refinement prompts that mix PII-containing and PII-free instances.
For every prompt $x \in B_t$, the loop performs the following steps.
\begin{enumerate}
\item The local SLM runs the full pipeline using the current skills and produces a response $y_s$ and a trace $\rho_s$ of the output of each stage.
\item The cloud LLM in its reference role answers the original prompt directly and produces a reference response $y_r$.
\item The cloud LLM in its evaluator role scores the triple $(x, y_s, y_r)$ and returns a quality score $q(x) \in \{1,\ldots,10\}$ together with a short failure tag $g(x)$ that names the symptom of any failure.
\item If $q(x) \ge \tau$ and no annotated identifier of $x$ reaches the cloud, the prompt counts as a success and the tag is ignored.
Otherwise, the prompt counts as a failure, and a deterministic rule uses $g(x)$ and trace counts in $\rho_s$ to attribute the failure to a single stage.
A prompt whose evaluator reply cannot be parsed counts as a failure only if an annotated identifier of $x$ reaches the cloud.
\end{enumerate}

\paragraph{Skill update.}
After every prompt in $B_t$ has been scored, the loop aggregates the per-prompt failure tags and applies three rules in sequence to update at most one skill.

\textit{Stage assignment.} Each failed prompt $x$ is mapped to the stage $j(x)$, namely decomposition, paraphrasing, or reconstruction, that is most likely the cause of the failure.
(i) An empty or off-topic output and any other symptom that (ii)--(iv) do not cover are mapped to reconstruction.
(ii) An excessive sub-task count is mapped to decomposition.
(iii) A privacy or redaction tag is mapped to paraphrasing, and a residual identifier in the cloud-bound request to paraphrasing, or to decomposition if its sub-task was sent unchanged.
(iv) A merge or coherence symptom is mapped to reconstruction, and a split symptom to decomposition.
The PII-aware routing skill is not included in this mapping because no failure symptom in (i)--(iv) is specific to routing.
The number of failures linked to stage $j$ in iteration $t$ is
\begin{equation}
c_j = \big|\{x \in B_t : x \text{ fails},\; j(x) = j\}\big|,
\end{equation}
and the stage $j^\star = \arg\max_j c_j$ performs worst.
If $c_{j^\star} = 0$, the iteration ends without an update.

\textit{Rewrite by the supervisor.} The cloud LLM in its supervisor role receives only aggregate failure statistics of the recent mini-batches, not the failing samples themselves, and returns a candidate rewrite of the current skill prompt for stage $j^\star$.
The aggregate-only input discourages the rewrite from copying the surface form of any single sample, and the supervisor is instructed to express the change as abstract behavioral guidance.

\textit{Memorization check.} Before a candidate is adopted, the check rejects placeholder slot tokens such as \texttt{[NAME]}, \texttt{<PII>}, or \texttt{[REDACTED]}, because such tokens would tie the skill to a specific entity category rather than encode a general detection rule.
Long quoted text segments and any eight-word sequence copied from a refinement or validation prompt are rejected, because such copies signal memorization of a specific sample.
Dataset names, benchmark-domain terms, and annotated identifiers of the evaluation benchmarks are rejected.
A rejected candidate is discarded without a retry.
An accepted candidate replaces the previous version of the skill prompt for stage $j^\star$, and each version is stored for reproducibility.
After each round, the loop keeps a revision only when its skill set lowers the leakage on the validation split and keeps side effects, such as the share of prompts answered without a cloud call and the retention of PII-free entities, within set bounds.

\paragraph{Stage-wise tuning.}
A stage-wise procedure outside the loop tunes the routing, decomposition, and paraphrasing skills.
For one stage at a time, the supervisor rewrites the skill from aggregate statistics of that stage on the refinement prompts, and a candidate that passes the memorization check is adopted only if it improves a stage-level metric, such as the share of annotated identifiers that the stage leaves exposed.
The stage-wise tuning chooses the version of each stage on the validation split by the bounds on side effects of that stage and then by its stage-level metric.
Because the validation runs of either procedure do not execute the reconstruction skill, a reconstruction revision is kept only as part of a skill set that meets the validation rule of the loop.
The final skill set of each SLM is selected at the lowest validation leakage within the bounds on side effects.
It keeps a second initial version, written in the same way after an analysis of the training and validation results, of the routing, paraphrasing, and reconstruction skills on Gemma4, the reconstruction skill on Llama3.2, and the paraphrasing skill on Qwen3.5.

\section{Experiments}
\label{sec:experiments}

\subsection{Experimental Setup}
\label{sec:setup}

\paragraph{Datasets.}
PRISM-160 comprises the 160 prompts of the PRISM benchmark~\citep{zhan2025prismprivacyawareroutingadaptive}, and this paper annotates 87 names and last four card digits in its 80 Medical and Banking prompts as identifiers by regular expressions for the PRISM prompt templates.
The PUPA test split comprises the 600 WildChat~\citep{zhao2024wildchat} user queries of PUPA~\citep{siyan2025papillon} that have listed PII units and lie outside the PUPA-New rows below 300 on which PAPILLON is optimized, and the 1,463 listed PII units that occur in 581 of these queries are the annotated identifiers.
The training split comprises 160 PUPA-New queries below row 300, 10 AI4Privacy~\citep{ai4privacy2024pii} and 10 Nemotron-PII~\citep{nvidia2025nemotronpii} instances used without their PII annotations, and 160 HotpotQA~\citep{yang2018hotpotqa} questions, and the validation split comprises 137 other PUPA-New queries below row 300 and 50 HotpotQA questions.
Neither split contains a PRISM-160 prompt, but the decomposition skills descend from versions refined on an earlier split that contained 40 PRISM-160 prompts, 100 PUPA test queries repeat the text of queries of the two splits, and 23 validation queries repeat the text of training queries.
P2Skill refines its skills on the training split, selects them on the validation split, and is tested on PRISM-160 and the PUPA test split.

\begin{table*}[tp]
\centering
\scriptsize
\setlength{\tabcolsep}{6pt}
\resizebox{\textwidth}{!}{%
\begin{tabular}{@{}ll ccccc ccccc@{}}
\toprule
& & \multicolumn{5}{c}{\textbf{PRISM-160}} & \multicolumn{5}{c}{\textbf{PUPA test}} \\
\cmidrule(lr){3-7} \cmidrule(lr){8-12}
\textbf{Local SLM} & \textbf{Method} & Leak$\downarrow$ & J-Leak$\downarrow$ & Q$\uparrow$ & IQ$\uparrow$ & \textbf{PQ}$\uparrow$ & Leak$\downarrow$ & J-Leak$\downarrow$ & Q$\uparrow$ & IQ$\uparrow$ & \textbf{PQ}$\uparrow$ \\
\midrule
\multirow{3}{*}{--} & Cloud Only                             & 100.0 & 98.9 & -- & 9.06 & 0.00 & 100.0 & 97.0 & -- & 8.66 & 0.00 \\
                    & Presidio                           & 37.9 & 37.9 & 0.344 & 8.04 & 5.03 & 46.4 & 49.4 & 0.528 & 7.63 & 4.00 \\
                    & Presidio+GLiNER                    & 2.3 & 2.3 & 0.325 & 8.01 & \textbf{8.19} & 22.8 & 24.4 & 0.452 & 7.42 & \textbf{5.61} \\
\midrule
\multirow{6}{*}{Gemma4} & Local Only                         & 0.0 & -- & 0.938 & 9.11 & -- & 0.0 & -- & 0.877 & 8.64 & -- \\
                           & Uniform LDP                        & 69.3 & 66.7 & 0.731 & 8.29 & 2.81 & 66.9 & 73.6 & 0.567 & 7.71 & 2.54 \\
                           & Selective LDP                      & 54.0 & 54.0 & 0.725 & 8.61 & 4.45 & 99.0 & 96.0 & 0.622 & 7.98 & 0.11 \\
                           & PRISM                              & 63.2 & 62.1 & 0.131 & 5.88 & 1.89 & 28.8 & 28.1 & 0.285 & 5.35 & 2.70 \\
                           & PAPILLON                           & 87.4 & 77.0 & 0.950 & 9.05 & 1.06 & 78.8 & 71.3 & 0.870 & 8.61 & 1.21 \\
                           & P2Skill                            & 0.0 & 0.0 & 0.725 & 8.82 & \textbf{9.11} & 15.6 & 15.9 & 0.690 & 7.55 & \textbf{5.61} \\
\midrule
\multirow{6}{*}{Qwen3.5} & Local Only                         & 0.0 & -- & 0.800 & 7.86 & -- & 0.0 & -- & 0.843 & 7.97 & -- \\
                           & Uniform LDP                        & 69.3 & 66.7 & 0.375 & 7.36 & 2.45 & 66.9 & 73.6 & 0.322 & 6.21 & 1.95 \\
                           & Selective LDP                      & 54.0 & 54.0 & 0.306 & 7.81 & 3.92 & 99.0 & 96.0 & 0.355 & 6.91 & 0.09 \\
                           & PRISM                              & 63.2 & 62.1 & 0.394 & 5.86 & 2.48 & 28.8 & 28.1 & 0.320 & 5.16 & 2.67 \\
                           & PAPILLON                           & 88.5 & 87.4 & 0.875 & 8.58 & 0.78 & 59.1 & 61.0 & 0.685 & 7.73 & 2.28 \\
                           & P2Skill                            & 10.0 & 10.3 & 0.669 & 7.73 & \textbf{7.05} & 27.3 & 29.7 & 0.682 & 7.15 & \textbf{4.53} \\
\midrule
\multirow{6}{*}{Llama3.2} & Local Only                         & 0.0 & -- & 0.719 & 7.87 & -- & 0.0 & -- & 0.657 & 7.33 & -- \\
                           & Uniform LDP                        & 69.3 & 66.7 & 0.375 & 7.97 & 2.46 & 66.9 & 73.6 & 0.322 & 6.26 & 2.03 \\
                           & Selective LDP                      & 54.0 & 54.0 & 0.406 & 8.42 & 3.99 & 99.0 & 96.0 & 0.342 & 6.84 & 0.08 \\
                           & PRISM                              & 63.2 & 62.1 & 0.188 & 6.29 & 2.14 & 28.8 & 28.1 & 0.262 & 5.43 & 2.78 \\
                           & PAPILLON                           & 2.3 & 2.3 & 0.325 & 7.17 & \textbf{7.62} & 26.3 & 28.8 & 0.242 & 5.45 & 3.31 \\
                           & P2Skill                            & 66.3 & 67.8 & 0.713 & 8.57 & 2.06 & 28.3 & 29.3 & 0.602 & 7.10 & \textbf{3.90} \\
\bottomrule
\end{tabular}%
}
\caption{Leakage and quality on PRISM-160 and the PUPA test split.}
\label{tab:main}
\end{table*}

\paragraph{Models.}
Ollama\footnote{\url{https://ollama.com/library}} serves the local SLMs Gemma4:e2B~\citep{gemma4} and Qwen3.5:2B~\citep{qwen2026qwen35} at bf16 and Llama3.2:3B~\citep{meta2024llama32, llama32} at fp16 precision, and the thinking mode of Gemma4:e2B and Qwen3.5:2B is disabled.
GPT-4o~\citep{hurst2024gpt4o} is the cloud LLM of every method that calls one and, at temperature 0, the LLM judge of every metric that uses one, and Claude Sonnet, called by a model alias of Claude Code~\citep{anthropic2026claudecode}, is the cloud LLM of the skill refinement in Section~\ref{sec:distill}.

\paragraph{Compared methods.}
Cloud Only sends the original prompt to the cloud LLM, and Local Only answers it on the local SLM, each using the same instruction as the corresponding requests of P2Skill.
Uniform LDP and Selective LDP reimplement the two LDP baselines of PRISM~\citep{zhan2025prismprivacyawareroutingadaptive} by random word replacement and regular-expression matching instead of Laplace noise and named entity recognition, and Presidio~\citep{presidio} and Presidio+GLiNER~\citep{zaratiana2024gliner, nvidia2025glinerpii} replace detected spans by placeholders that the device restores in the cloud response.
PRISM~\citep{zhan2025prismprivacyawareroutingadaptive} and PAPILLON~\citep{siyan2025papillon} run from their official code, and PAPILLON uses its released prompt on Llama3.2 and a prompt re-optimized by its official script on the other two SLMs.

\paragraph{Metrics.}
String-match leakage (Leak) is the percentage of annotated identifiers that appear as a case-insensitive substring of any cloud-bound request, and for P2Skill, PAPILLON, and the LDP baselines it is the mean of three runs, of which only the first sends its requests to the cloud LLM, and the other metrics use this first run.
LLM-judged leakage (J-Leak) applies the leakage judge of PAPILLON~\citep{siyan2025papillon} to the same identifiers, and PAPILLON reports this judged metric as its leakage.
Quality preservation (Q) is the share of prompts on which the response does not lose to the target response, here the Cloud Only response, in the pairwise LLM judge of PAPILLON~\citep{siyan2025papillon}, and a failed or empty response counts as a loss.
Inference quality (IQ) is the rating of the response from 1 to 10 by the single-answer grading of MT-Bench~\citep{zheng2023judging}, as in PRISM~\citep{zhan2025prismprivacyawareroutingadaptive}.
Privacy preservation of a prompt is $\text{PP} = 1 - N_{\text{leakage}} / N_{\text{total}}$, where $N_{\text{leakage}}$ is the number of its annotated identifiers that the first run transmits and $N_{\text{total}}$ the number of its annotated identifiers.
Table~\ref{tab:main} gives, for each method that sends text to the cloud, the mean privacy-quality score $\text{PQ} = \text{IQ} \times \text{PP}$ of the prompts that contain annotated identifiers, and PQ does not apply to Local Only, which transmits nothing.
Bold marks the highest PQ of each block of rows in Table~\ref{tab:main}, and a dash marks a value that does not apply or is not computed.

\paragraph{Statistical analysis.}
The metrics, test splits, hypotheses, and tests are pre-specified in a written evaluation plan (provided as supplementary material) before the final GPT-4o evaluation, except that the PQ columns of Table~\ref{tab:main} are exploratory additions made after the results had been seen.
On PUPA, the three confirmatory hypotheses state that P2Skill on Gemma4, chosen for its lowest validation leakage, preserves quality more often than (H1) Presidio+GLiNER, (H2) PAPILLON on Llama3.2, the only SLM for which PAPILLON releases a prompt, and (H3) Local Only on Gemma4, and exact McNemar tests of the paired per-prompt Q outcomes test them at $\alpha = 0.05$ after the Holm correction.
Every comparison other than H1 to H3 is descriptive, that is, it gets a 95\% paired bootstrap confidence interval of 10,000 prompt resamples and no hypothesis test, and a method dominates another in the Pareto sense if it is better in Leak or Q by an interval that excludes zero and not worse in the other as a point estimate.

\subsection{Privacy Leakage and Inference Quality}
\label{sec:results}

On PUPA, P2Skill on Gemma4:e2B transmits 15.6\% of the annotated identifiers, fewer than every baseline that sends text to the cloud.
On this benchmark, P2Skill dominates the LDP baselines, PRISM, and Presidio on each SLM and Presidio+GLiNER on Gemma4, no baseline that sends text to the cloud dominates P2Skill on the same SLM, and P2Skill reaches a higher PQ than the LDP baselines, PRISM, and the pre-specified PAPILLON configuration on each SLM (Table~\ref{tab:main}).
Presidio+GLiNER reaches a higher PQ than P2Skill on Qwen3.5 and Llama3.2 on both benchmarks and the same PUPA PQ of 5.61 on Gemma4.
P2Skill preserves quality more often than Presidio+GLiNER on every SLM and both benchmarks, significantly in the confirmatory PUPA test on Gemma4 (H1).

In the second confirmatory test, P2Skill on Gemma4 preserves quality significantly more often than PAPILLON on Llama3.2 (H2), the PAPILLON configuration that has the lowest Q in Table~\ref{tab:main}, whereas PAPILLON on Gemma4 preserves quality more often than P2Skill but transmits more identifiers.
H3 fails, because Local Only on Gemma4 preserves quality more often than P2Skill and the test rejects equality in this opposite direction.
Following the rule of the evaluation plan for a failed H3, this paper makes no utility claim relative to Local Only, which dominates P2Skill on every SLM and both benchmarks.
An exploratory route-only variant of P2Skill answers a prompt on the local SLM if P2Skill sets $p_i{=}1$ for any of its sub-tasks or fails to decompose it and otherwise sends it unchanged to the cloud LLM.
The variant reaches a higher PUPA PQ than P2Skill on Gemma4 and Llama3.2, the two SLMs on which its validation Q and IQ qualified it for testing, which suggests that the paraphrase and reconstruction path of P2Skill loses information, although the variant routes whole prompts instead of sub-tasks.

On PRISM-160, the PQ of P2Skill depends on the SLM, from 9.11 on Gemma4, where P2Skill transmits none of the annotated identifiers in any run, to 2.06 on Llama3.2, where it transmits 66.3\% of them and every baseline that sends text to the cloud except Cloud Only has a higher PQ point estimate.
On Gemma4, the exploratory PAPILLON variant that uses its released prompt dominates P2Skill on PRISM-160.

\subsection{Routing Behavior and Privacy Analysis}
\label{sec:routing-behavior}
\begin{table}[t]
\centering
\footnotesize
\setlength{\tabcolsep}{2.5pt}
\resizebox{\columnwidth}{!}{%
\begin{tabular}{@{}ll r rr rr r@{}}
\toprule
& & & \multicolumn{2}{c}{\textbf{cloud}} & \multicolumn{2}{c}{\textbf{local}} & \\
\cmidrule(lr){4-5} \cmidrule(lr){6-7}
\textbf{Prompts} & \textbf{SLM} & $n$ & direct & \shortstack{para-\\phrased} & solvable & fallback & skipped \\
\midrule
\multirow{3}{*}{PII} & Gemma4 & 80 & 0.0 & 77.5 & 22.5 & 0.0 & 0.0 \\
 & Qwen3.5 & 97 & 12.4 & 43.3 & 44.3 & 0.0 & 0.0 \\
 & Llama3.2 & 80 & 43.8 & 51.2 & 5.0 & 0.0 & 0.0 \\
\midrule
\multirow{3}{*}{PII-free} & Gemma4 & 80 & 27.5 & 66.2 & 6.2 & 0.0 & 0.0 \\
 & Qwen3.5 & 140 & 37.1 & 14.3 & 48.6 & 0.0 & 0.0 \\
 & Llama3.2 & 86 & 88.4 & 5.8 & 5.8 & 0.0 & 0.0 \\
\midrule
\multirow{3}{*}{PUPA} & Gemma4 & 621 & 20.9 & 38.6 & 28.7 & 1.3 & 10.5 \\
 & Qwen3.5 & 1688 & 36.6 & 30.2 & 32.2 & 0.9 & 0.1 \\
 & Llama3.2 & 709 & 36.7 & 44.0 & 12.1 & 2.3 & 4.9 \\
\bottomrule
\end{tabular}}
\caption{Sub-task routes of P2Skill (\% of the $n$ sub-tasks).
PII denotes the Medical and Banking prompts of PRISM-160, PII-free its other prompts, and fallback a paraphrase that fails the identifier matcher.}
\label{tab:routing-pii}
\end{table}

Table~\ref{tab:routing-pii} gives the routes of the sub-tasks of P2Skill in the first run, and the identifiers that P2Skill transmits reach the cloud in sub-tasks that have $p_i{=}0$ and in paraphrases that keep an identifier.
On PRISM-160, every patient name that P2Skill on Llama3.2 leaks in the three runs reaches the cloud in a sub-task that it sends unchanged.
On PUPA, sub-tasks that have $p_i{=}0$ carry most of the transmitted identifiers on Gemma4 and Llama3.2, and paraphrases carry most on Qwen3.5.
The identifier matcher recognizes none of the PRISM-160 identifiers, which leaves their detection to the decomposition skill, and the density threshold keeps no sub-task of the first run local.

\FloatBarrier
\section{Conclusion}
\label{sec:conclusion}

This paper proposes P2Skill, a prompt-based skill distillation method for cloud-local LLM inference, in which a frozen local small language model (SLM) executes four privacy-preserving skills and a cloud LLM guides their refinement, without any model training.
On real user queries, P2Skill on its best local SLM transmits fewer annotated identifiers than every existing privacy-preserving method that sends text to the cloud and preserves quality more often than the closest of them in leakage.
Fully local answering preserves quality more often than P2Skill on these queries, and the leakage of P2Skill depends on the local SLM.

\section*{Limitations}

P2Skill provides no formal privacy guarantee, and its leakage is measured only on the annotated identifiers of each benchmark, which leaves identifying details outside these identifiers uncounted.
The leakage and quality of P2Skill depend on the local SLM, and answering every real user query on the local SLM alone preserves quality more often than P2Skill.
The experiments do not separate the effect of the refinement loop and the stage-wise tuning from that of the second initial skill versions, which three of the four final skills on each of Gemma4 and Qwen3.5 and two of the four on Llama3.2 keep or descend from.
Quality is measured by LLM judges~\citep{zheng2023judging, li2024llmsasjudges, wu2025style, li2025exploring}, which may not reflect human preferences on the privacy-utility trade-off.
The evaluation covers two mostly English benchmarks and a limited set of domains, and it does not test P2Skill on requests selected to exceed the capability of the local SLM.

\section*{Ethical Considerations}

PUPA builds on WildChat queries of users who opted in to the collection of their conversations~\citep{zhao2024wildchat} and contains personal information of real users, and this paper shows no PUPA text.
The prompt examples of this paper are either constructed for illustration or synthetic PRISM-160 prompts.
As in the PAPILLON protocol, the experiments send PUPA queries to GPT-4o and GPT-4o-mini by the OpenAI API, and the skill refinement and a second LLM judge send them to Claude Sonnet by the Claude Code command-line interface, according to the terms of each provider.
P2Skill is intended for research on privacy protection, gives no formal privacy guarantee, and should not be the only safeguard for personal data.

\bibliography{references}

\appendix
\section{Evaluation Details and Disclosures}
\label{app:disclosures}

The evaluation plan that this appendix cites, including its dated amendments, is part of the supplementary material.

\subsection{Skill and Prompt Examples}
\label{app:examples}

\paragraph{Skill excerpts.}
\label{app:skills}
The boxes give verbatim excerpts, chosen for illustration, of the frozen skills of P2Skill on Gemma4:e2B, and \texttt{[...]} marks a cut.
On this SLM, only the decomposition skill results from the iterative refinement loop and the stage-wise tuning, and the other three skills are second initial versions.
The refinement replaces the short identifier list of the initial decomposition skill by a procedure that lists every private fact of a sub-task and sets the PII flag if it finds one or if the SLM is unsure.

\par\vspace{3pt}
\begingroup
\providecommand{\PSkCut}{\par[...]}
\providecommand{\PSkCutIn}{[...]}
\providecommand{\PSkUser}{\par{\rmfamily\itshape User prompt template}}
\providecommand{\PSkItem}{\hangindent=1.2em\hangafter=1 }
\setlength{\fboxsep}{2pt}

\par\noindent\fbox{\begin{minipage}{\dimexpr\columnwidth-2\fboxsep-2\fboxrule\relax}
\scriptsize\ttfamily\raggedright\setlength{\parskip}{0.8pt}\setlength{\parindent}{0pt}
{\rmfamily\bfseries Decomposition skill (\#1), initial version}
\par You are a task decomposition expert. Your job is to split user prompts ONLY when the prompt contains multiple distinct requests or tasks.
\PSkCut{}
\par \PSkItem 3. When in doubt, keep it as 1 subtask. \PSkCutIn{}
\par \PSkItem 4. Mark any subtask containing personal information (names, emails, phones, addresses, IDs, etc.) with has\_pii: true \PSkCutIn{}
\end{minipage}}
\par\vspace{3pt}

\par\noindent\fbox{\begin{minipage}{\dimexpr\columnwidth-2\fboxsep-2\fboxrule\relax}
\scriptsize\ttfamily\raggedright\setlength{\parskip}{0.8pt}\setlength{\parindent}{0pt}
{\rmfamily\bfseries Decomposition skill (\#1), final version}
\par You are a task decomposition engine for a privacy pipeline.
\par \PSkCutIn{} Always decompose it. Never refuse and never output an empty array, even for sensitive or personal content. \PSkCutIn{}
\par \PSkItem 1. SPLIT. Use one subtask for the whole input by default. \PSkCutIn{} Never drop, shorten or rephrase anything.
\par \PSkItem 2. LIST. For each subtask, on its own, read it once from start to end and list every private fact about a real person. Private facts are: names, nicknames, initials, \PSkCutIn{} family or other relationships, dates tied to a person, and health, money, legal or personal-life facts. Also count facts that are only implied by the requester\textquotesingle{}s own situation, \PSkCutIn{} The requester counts as a person whenever the text says I, me, my or we.
\par \PSkItem 3. DECIDE. If the list has even one item, set has\_pii=true \PSkCutIn{} When unsure, choose has\_pii=true.
\par \PSkItem 4. CHECK. Make sure the JSON is valid \PSkCutIn{}
\end{minipage}}
\par\vspace{3pt}

\par\noindent\fbox{\begin{minipage}{\dimexpr\columnwidth-2\fboxsep-2\fboxrule\relax}
\scriptsize\ttfamily\raggedright\setlength{\parskip}{0.8pt}\setlength{\parindent}{0pt}
{\rmfamily\bfseries Routing skill (\#2), final version}
\par \PSkCutIn{} Classify each task into exactly one label:
\par \PSkItem - no\_answer\_needed: this subtask is context, instruction, or metadata that doesn\textquotesingle{}t require a separate answer \PSkCutIn{}
\par \PSkItem - slm\_solvable: a small local model can complete it reliably \PSkCutIn{}
\par \PSkItem - slm\_unsolvable: requires broader knowledge, complex reasoning, \PSkCutIn{}
\par \PSkItem - needs\_decomposition: task is too complex \PSkCutIn{}
\PSkUser{}
\par Task: \{task\_content\}
\par Has PII: \{has\_pii\} \PSkCutIn{}
\end{minipage}}
\par\vspace{3pt}

\par\noindent\fbox{\begin{minipage}{\dimexpr\columnwidth-2\fboxsep-2\fboxrule\relax}
\scriptsize\ttfamily\raggedright\setlength{\parskip}{0.8pt}\setlength{\parindent}{0pt}
{\rmfamily\bfseries Paraphrasing skill (\#3), final version}
\par Rewrite the request so that it contains NO detail that identifies the user or any specific person, organization or place they deal with. Keep everything else the same.
\par Replace each identifying detail with a short generic description:
\par \PSkItem - a person\textquotesingle{}s name becomes their role, for example the applicant, my manager, a colleague, a friend
\par \PSkItem - a company, employer, school, \PSkCutIn{} name becomes its type, for example a university, a software firm, the clinic
\PSkCut{}
\par A detail that appears several times must be replaced every time, including short forms such as a first name alone or initials. \PSkCutIn{}
\end{minipage}}
\par\vspace{3pt}

\par\noindent\fbox{\begin{minipage}{\dimexpr\columnwidth-2\fboxsep-2\fboxrule\relax}
\scriptsize\ttfamily\raggedright\setlength{\parskip}{0.8pt}\setlength{\parindent}{0pt}
{\rmfamily\bfseries Reconstruction skill (\#4), final version}
\par \PSkCutIn{} The partial answers were written without the user\textquotesingle{}s personal details, so they may use generic descriptions such as the applicant, a company or a university. \PSkCutIn{}
\par \PSkItem 2. Wherever a partial answer uses a generic description, put back the specific detail from the original request, such as the real names, organizations, places or numbers, \PSkCutIn{}
\par \PSkItem 3. Use only facts from the partial answers and the original request. \PSkCutIn{}
\end{minipage}}
\par\vspace{3pt}

\endgroup

\paragraph{Prompt examples.}
\label{app:prompt-examples}
Table~\ref{tab:prompt-examples} shows the records of P2Skill for three synthetic PRISM-160 prompts, and no PUPA text is shown because PUPA contains personal information of real users.
Each prompt is the first draw of the fixed sampler of the evaluation plan in its stratum.
The three strata, which the plan does not fix, are the largest stratum of prompts whose identifier stays on the device and the two strata of prompts whose identifier reaches the cloud on Llama3.2:3B, the SLM that leaks most on PRISM-160.
Bold marks the annotated identifier, Leak gives the number of transmitted annotated identifiers in each of the three runs, Q gives the outcome of the pairwise judgment against Cloud Only, and Q, IQ, and the texts come from the first run.
In (a), the paraphrasing skill removes the name and the reconstruction skill restores it on the device, in (b), neither the decomposition skill nor the identifier matcher flags the name, and in (c), the paraphrase keeps the last four card digits, which the matcher does not recognize.
Prompt B25 of (c) belongs to the earlier training split that contained 40 PRISM-160 prompts.

\begin{table}[tp]
\centering
\footnotesize
\setlength{\tabcolsep}{3pt}
\renewcommand{\arraystretch}{1.05}
\begin{tabular}{@{}l p{0.80\columnwidth}@{}}
\toprule
\multicolumn{2}{@{}l@{}}{\textbf{(a) M11}, Gemma4:e2B, no leak, paraphrased (62 prompts)} \\
\midrule
Prompt & \raggedright A 7-year-old female patient named \textbf{Mia} reports symptoms: pain, fatigue, indigestion, bloating. What is the likely diagnosis? \tabularnewline
Sub-task & \raggedright The whole prompt, $p_1{=}1$, cloud paraphrased \tabularnewline
To cloud & \raggedright A 7-year-old female patient reports symptoms: pain, fatigue, indigestion, bloating. What is the likely diagnosis? \tabularnewline
Answer & \raggedright When evaluating a 7-year-old female patient named \textbf{Mia} [...], one likely diagnosis [...] is celiac disease. [...] \tabularnewline
Result & \raggedright Leak 0/0/0, Q tie, IQ 9 \tabularnewline
P+G & \raggedright A [AGE\_1] female patient named [PERSON\_1] reports symptoms: [...] (Leak 0, Q loss, IQ 7) \tabularnewline
\midrule
\multicolumn{2}{@{}l@{}}{\textbf{(b) M35}, Llama3.2:3B, leaked, direct (35 prompts)} \\
\midrule
Prompt & \raggedright A 31-year-old female patient named \textbf{Lily} reports symptoms: cramping, altered\_appetite, pain, altered\_weight, stiffness, fever. What is the likely diagnosis? \tabularnewline
Sub-task & \raggedright The whole prompt in \texttt{<input>} tags, $p_1{=}0$, cloud direct \tabularnewline
To cloud & \raggedright \texttt{<input>}A 31-year-old female patient named \textbf{Lily} reports symptoms: [...] What is the likely diagnosis?\texttt{</input>} \tabularnewline
Answer & \raggedright To provide a likely diagnosis for \textbf{Lily}, a 31-year-old female [...] Inflammatory Bowel Disease (IBD) [...] \tabularnewline
Result & \raggedright Leak 1/1/1, Q tie, IQ 9 \tabularnewline
P+G & \raggedright A [AGE\_1] [PERSON\_1] patient named [PERSON\_2] reports symptoms: [...] (Leak 0, Q loss, IQ 9) \tabularnewline
\midrule
\multicolumn{2}{@{}l@{}}{\textbf{(c) B25}, Llama3.2:3B, leaked, paraphrased (26 prompts)} \\
\midrule
Prompt & \raggedright This is Lily writing to report regarding a charge of \$34 on my PNC Bank card ending with \textbf{2691}, dated 2024-01-25 at a restaurant in Cleveland. Please help me write an email to submit this dispute properly. \tabularnewline
Sub-task & \raggedright The whole prompt in \texttt{<input>} tags, $p_1{=}1$, cloud paraphrased \tabularnewline
To cloud & \raggedright This is a customer writing to report regarding a charge of \$34 on their bank card ending with \textbf{2691}, dated 2024-01-25 at a restaurant in a major Midwestern city. [...] \tabularnewline
Answer & \raggedright [...] Dear PNC Bank Customer Service, [...] dispute a charge on my bank card ending in \textbf{2691}. [...] made at a restaurant located in Cleveland. [...] \tabularnewline
Result & \raggedright Leak 1/1/1, Q loss, IQ 9 \tabularnewline
P+G & \raggedright This is [PERSON\_1] writing to report regarding a charge of \$[AGE\_1] on my PNC Bank card ending with [US\_BANK\_NUMBER\_1], dated [DATE\_TIME\_1] at a restaurant in [LOCATION\_1]. [...] (Leak 0, Q loss, IQ 9) \tabularnewline
\bottomrule
\end{tabular}
\caption{P2Skill examples on PRISM-160, where P+G denotes the request of Presidio+GLiNER.}
\label{tab:prompt-examples}
\end{table}

\subsection{Data Use and Test-Set Access}
\label{app:data}

\paragraph{Annotation and data use.}
The regular expressions that annotate PRISM-160 miss names in other phrasings, and a heuristic count finds a customer first name in 31 of the 40 Banking prompts, whereas only 7 Banking names are annotated.
This omission makes the transmitted counts on PRISM-160 a lower bound for every method, whereas a leakage percentage can rise or fall once these names are counted.
In the three leaked Qwen3.5:2B prompts that the fixed sampler draws (B3, B22, and B26), the paraphrase of P2Skill keeps the last four card digits, and in B22 and B26 it repeats the first sentence of the prompt unchanged, including a customer first name that the annotation does not count.
A PUPA PII unit counts for its query only if it has at least two characters, is not a Presidio placeholder of the released PUPA data, and occurs in the query as a case-insensitive substring.
Skill refinement and evaluation use public benchmarks and collect no new user data, and Claude Sonnet, the cloud LLM of the refinement, receives the PUPA queries of the training split and, for the routing labels of the stage-wise tuning, the PII-flagged sub-tasks of the PUPA validation queries.

\paragraph{Overlaps and selection.}
The decomposition skill of every SLM descends from versions refined on an earlier training split that contained 40 PRISM-160 prompts, and none of the frozen skills descends from a later training split that contained 40 PRISM-160 prompts as well.
An AI assistant writes the initial and the second initial versions of every skill from human instructions, the second after analysis of results on the training and validation splits.
The second initial routing skill differs from the first only in one replaced example, the Llama3.2:3B and Qwen3.5:2B routing skills are tuned from second initial versions, and the Qwen3.5:2B reconstruction skill is refined from one.
The 100 PUPA test queries that repeat PUPA-New rows below 300 lie in the rows on which the released and re-optimized PAPILLON prompts are optimized, and the test split is used without deduplication, as in PAPILLON.
The final skills are selected at the lowest validation leakage within bounds on the share of prompts answered without a cloud call and on the retention of PII-free entities, both set after validation results had been seen, and the first bound was later raised from 0.35 to 0.40, a change that lets the frozen Gemma4:e2B skills meet it.

\paragraph{Test-set access.}
Before the evaluation plan was fixed, the leakage of P2Skill on 150 PUPA test queries and of the baselines on all PUPA test queries was measured, and after the plan was fixed, unit tests that replace the cloud LLM by a stub printed the leakage of the untuned PRISM on both test sets.
The PRISM-160 leakage of P2Skill on two SLMs and of several baselines was inspected before the skills of the third SLM were frozen, after which no skill or configuration pre-specified in the evaluation plan changed in response to test measurements, the Qwen3.5:2B skills were selected on the validation split only, and the exploratory variants were added after test results had been seen.

\paragraph{Artifacts and licenses.}
Table~\ref{tab:licenses} gives the license or terms of each dataset, model, and tool that this paper uses.
This paper uses every artifact for non-commercial research on privacy-preserving inference, a use that the license or terms of each artifact permit and that matches its intended use.

\begin{table}[tp]
\centering
\footnotesize
\begin{tabular}{@{}p{0.53\columnwidth} p{0.40\columnwidth}@{}}
\toprule
\textbf{Artifact} & \textbf{License or terms} \\
\midrule
\raggedright PUPA and PAPILLON code~\citep{siyan2025papillon} & \raggedright MIT \tabularnewline
\raggedright WildChat~\citep{zhao2024wildchat} & \raggedright ODC-BY 1.0 \tabularnewline
\raggedright PRISM-160 and PRISM code~\citep{zhan2025prismprivacyawareroutingadaptive} & \raggedright MIT \tabularnewline
\raggedright AI4Privacy pii-masking-200k~\citep{ai4privacy2024pii} & \raggedright AI4Privacy free license \tabularnewline
\raggedright Nemotron-PII~\citep{nvidia2025nemotronpii} & \raggedright CC BY 4.0 \tabularnewline
\raggedright HotpotQA~\citep{yang2018hotpotqa} & \raggedright CC BY-SA 4.0 \tabularnewline
\raggedright Gemma4:e2B~\citep{gemma4}, Qwen3.5:2B~\citep{qwen2026qwen35} & \raggedright Apache 2.0 \tabularnewline
\raggedright Llama3.2:3B~\citep{meta2024llama32} & \raggedright Llama 3.2 Community License \tabularnewline
\raggedright GPT-4o and GPT-4o-mini~\citep{hurst2024gpt4o} & \raggedright OpenAI terms and usage policies \tabularnewline
\raggedright Claude Sonnet and Claude Code~\citep{anthropic2026claudecode} & \raggedright Anthropic terms and usage policy \tabularnewline
\raggedright Presidio~\citep{presidio}, DSPy~\citep{khattab2024dspy}, Ollama & \raggedright MIT \tabularnewline
\raggedright GLiNER library~\citep{zaratiana2024gliner} & \raggedright Apache 2.0 \tabularnewline
\raggedright GLiNER-PII~\citep{nvidia2025glinerpii} & \raggedright NVIDIA Open Model License \tabularnewline
\raggedright MT-Bench judge prompts of FastChat~\citep{zheng2023judging} & \raggedright Apache 2.0 \tabularnewline
\bottomrule
\end{tabular}
\caption{Licenses and terms of the datasets, models, and tools.}
\label{tab:licenses}
\end{table}

\subsection{Adaptations of the Compared Methods}
\label{app:adaptations}

Every compared method uses the cloud LLM and the local SLMs of this evaluation, PAPILLON keeps its official limit of 4000 tokens for the cloud response, and PRISM keeps its official limit of 200 tokens for the cloud sketch, a draft response of the cloud LLM that the local SLM refines.

\paragraph{LDP baselines.}
Uniform LDP replaces each prompt word by a generic word at probability $1/(1+e^{\epsilon})$, where $\epsilon = 1$, and Selective LDP replaces each span that its regular expressions detect by a substitute from a fixed list, a step that ignores $\epsilon$.
Both reimplementations send the system prompt ``Answer this request:'' instead of the Cloud Only instruction, and their local refinement prompt keeps only the first 300 characters of the user prompt and the first 500 characters of the cloud response, which can lower their quality.
Selective LDP uses the patterns of the identifier matcher of P2Skill and the template that produced the 40 Medical names of PRISM-160, which makes its PRISM-160 leakage optimistic, although it transmits all 47 Banking identifiers.

\paragraph{Presidio and Presidio+GLiNER.}
Presidio uses its default recognizers and an anonymizer that inserts unnumbered type tags, and this paper adds numbered placeholders, their on-device map and restoration, and a system-prompt note that asks the cloud LLM to copy each placeholder exactly.
The detected spans are frozen once per dataset, because the detection of Presidio depends on the hash seed of Python.

\paragraph{PRISM.}
PRISM uses its released router checkpoint, its default privacy parameters $\epsilon = 2$, $\alpha = 0.5$, and $\lambda = 0.4$, and, as in its official code, a cloud sketch sampled at temperature 0.3 and top-p 0.9 that it returns as the answer on its cloud-only route.
The released router checkpoint of PRISM is trained on data that include the 40 Medical prompts of PRISM-160.
Ollama serves the edge model of PRISM for Gemma4:e2B and Qwen3.5:2B in raw completion mode, because the official llama.cpp engine loads only Llama3.2:3B of the three SLMs, and the official context window of 2048 tokens is emulated.
On PUPA, PRISM fails on 23, 20, and 22 queries on Gemma4:e2B, Qwen3.5:2B, and Llama3.2:3B, of which 18, 15, and 15 exceed this context window, which counts as a method failure as on the official engine, and the others return an empty answer.
This evaluation corrects the official loader of the edge few-shot examples, which drops the example responses, assigns the examples to wrong domains, and fails every edge-side call on Medical, Banking, and Tourism prompts.
PRISM reaches GPT-4o by the OpenAI API instead of its default third-party relay, and the evaluation prepends the beginning-of-sequence token to the raw prompt of its edge model on Gemma4:e2B and appends an empty thinking block to it on Qwen3.5:2B, as the Ollama chat templates of these models do in the other methods.

\paragraph{PAPILLON.}
PAPILLON uses its released prompt on Llama3.2:3B, and for Gemma4:e2B and Qwen3.5:2B, the official optimization script is rerun on the same SLM weights, keeping its MIPROv2 optimizer, the PUPA-New training rows 0--149 and validation rows 150--299, and GPT-4o-mini (\texttt{gpt-4o-mini-2024-07-18}) as prompt model and judge.
Both the released and the re-optimized prompts are optimized for GPT-4o-mini as the cloud LLM, whereas this evaluation uses GPT-4o.
The optimizer comes from DSPy 2.4.16, the only release that provides both the client and the optimizer interface of the script, and runs on the DSPy 2.5.35 runtime that evaluates every PAPILLON prompt, whereas the released Llama3.2:3B prompt was most likely optimized in the prompt rendering of DSPy 2.4 and is evaluated in that of DSPy 2.5.
The leakage judge of this script counts a leak only if the reply of GPT-4o-mini has the prefix ``yes'', and a reply that first repeats the judge template counts as no leak whatever its verdict, a parsing that is identical in DSPy 2.4.16 and 2.5.35 and that this evaluation does not change.
In the re-optimization, 12,856 of the 19,366 judge replies for Gemma4:e2B and 7,873 of the 12,922 for Qwen3.5:2B that start by repeating the template contain a yes verdict that the parser does not read.
The prompt selected for Gemma4:e2B scores 89.13 on validation by the official metric but 63.1 when these verdicts are read, and it already copied about 75\% of the PII units into the cloud request during optimization, which indicates that its test leakage comes from the selected prompt and not from the evaluation setup.
The re-optimized prompts stay the pre-specified configuration, because a modified procedure would no longer reproduce PAPILLON, and an exploratory variant, added after this finding, runs the released Llama3.2:3B prompt unchanged on Gemma4:e2B and Qwen3.5:2B.
PAPILLON runs the three stages of its official forward pass one at a time, which lets the evaluation retry infrastructure errors that the official code turns into an empty answer.

\subsection{Runs, Judges, and Additional Analyses}
\label{app:repro}

\paragraph{Refinement settings.}
The refinement loop leaves out 150 of the HotpotQA questions of the training split, and the stage-wise tuning leaves out its AI4Privacy and Nemotron-PII instances.
The refinement loop of Section~\ref{sec:distill} uses $\tau = 7$, rounds of $T = 8$ iterations, and mini-batches of $|B_t| = 5$ prompts, and the supervisor receives the failure counts and the mean score of the last mini-batch and, in the runs on the training split, the identifier-level counts of the last three mini-batches.
A run of the loop has at most five rounds.
In the rounds that produced the final decomposition and reconstruction skills of Qwen3.5, a revision of the decomposition or paraphrasing skill is rolled back if its leakage on a fixed subset of the refinement prompts rises by more than two percentage points or if a bound on side effects, such as at most 20\% skipped sub-tasks, fails.
In the runs that continue an earlier run, a round counts as an improvement when it lowers the validation leakage by more than 1 or 1.5 percentage points, and the run stops after two or three rounds without improvement.
In each of at most five rounds, the stage-wise tuning scores three candidate rewrites and adopts the best one if it improves the stage-level metric by more than 0.01, and the validation split then chooses between the starting version and the last two adopted versions.
The final selection requires at most 40\% of the PUPA validation queries without a cloud call, at least 60\% retention of PII-free entities, at most 10\% paraphrases that answer the request instead of rewriting it, and at most 20\% skipped sub-tasks.

\paragraph{Runs and decoding.}
Every GPT-4o request uses the pinned snapshot \texttt{gpt-4o-2024-\allowbreak 08-\allowbreak 06}, which is verified on every response, and P2Skill, Cloud Only, and the Presidio and LDP baselines query it at temperature 0 and at most 1024 output tokens.
Ollama serves the SLMs on NVIDIA L4 GPUs, and P2Skill and Local Only decode at temperature 0.3 and at most 2048 output tokens, except for the routing and paraphrasing skills at temperature 0.2 and at most 384 and 1024 output tokens.
Local decoding on a shared Ollama server is not bit-reproducible because its prefix cache depends on earlier requests, and the pooled leakage of a method may mix servers that process four requests in parallel or one at a time.
The route shares of P2Skill in Table~\ref{tab:routing-pii} come from the first run, and those of the second and third runs differ from them by at most 10 percentage points.
For P2Skill and PAPILLON on Llama3.2:3B on PUPA, complete runs replace interrupted second runs of 221 and 121 records from a server of another configuration, and each pooled run is complete and uses one server configuration.
The LDP noise, unseeded in the original code of PRISM, is seeded per prompt and run, which gives an LDP baseline the same cloud-bound text on the three SLMs.

\paragraph{Compute.}
The local SLMs run on two NVIDIA L4 GPUs of 24~GB each.
The wall times of all prompt records sum to about 404 hours for the final evaluation, validation runs included, and to about 235 hours for the 13,680 prompt records of the transfer runs.
This sum overstates the GPU time because it includes the latency of cloud calls and several runs share a GPU, and it excludes the skill refinement and tuning.
The final evaluation makes 75,779 GPT-4o calls, judges included, for 295.60 USD, and the PAPILLON re-optimization makes 42,608 GPT-4o-mini calls for 12.17 USD.
Claude Sonnet runs by the Claude Code command-line interface on a subscription.
The skill refinement records no per-call cost, and for the 9,927 calls of the second judge the interface reports an API-equivalent cost of about 171 USD, which the subscription does not bill separately.

\paragraph{Judges.}
The IQ judge follows the single-answer grading of MT-Bench, except that GPT-4o replaces GPT-4, each answer is judged in one turn, and no reference answer is used, as in MT-Bench outside its math, reasoning, and coding categories.
A prompt counts as a loss in Q when the final response is judged worse than the Cloud Only response in one presentation order and not better in the other.
A failed or empty response counts as a loss in Q without a judge call, and IQ grades the returned response as it is.
PAPILLON fails on 5 PUPA queries in its re-optimized configuration on Qwen3.5:2B and on 1 and 3 PUPA queries in its released-prompt configuration on Gemma4:e2B and Qwen3.5:2B, and PRISM fails on one PRISM-160 prompt on Qwen3.5:2B.
No GPT-4o judge call fails or is refused, and the one unparseable IQ judgment, of PRISM on Gemma4:e2B on PUPA, is dropped from the mean and counted, as in the MT-Bench code.

\paragraph{Failure cases of P2Skill.}
P2Skill returns an empty response when every sub-task of a prompt is skipped, which happens on 60, 1, and 13 PUPA queries on Gemma4:e2B, Qwen3.5:2B, and Llama3.2:3B and on no PRISM-160 prompt.
In the 60 Gemma4:e2B queries, the decomposition skill drops the instruction and keeps the content, the routing skill labels the remaining sub-tasks as needing no answer, and the pipeline has no fallback, whereas Local Only answers all 60.
The sensitivity check of the evaluation plan, which substitutes the Local Only responses of the same SLM for the empty responses, raises the Q of P2Skill to 0.785, 0.683, and 0.617 and leaves the verdicts of H1 to H3 unchanged (H3 then has 36 against 91 discordant queries, $p = 1.1 \times 10^{-6}$).
The Q difference between P2Skill and Local Only is then $-0.092$ [$-0.128$, $-0.055$], $-0.160$ [$-0.202$, $-0.120$], and $-0.040$ [$-0.088$, 0.010] on Gemma4:e2B, Qwen3.5:2B, and Llama3.2:3B.
P2Skill on Qwen3.5:2B sends copies of its decomposition-skill instructions to the cloud LLM as sub-tasks in at least 4 of its 126 PRISM-160 and 34 of its 1,127 PUPA cloud requests of the first run, and these copies contain no user identifier.
Gemma4:e2B and Qwen3.5:2B run the same second initial paraphrasing skill, and in the first PUPA run, 242 of the 430 annotated identifiers that Qwen3.5:2B transmits reach the cloud only in its 509 paraphrases, against 63 of 230 in the 240 paraphrases of Gemma4:e2B.
Qwen3.5:2B splits the 600 queries into 1,688 sub-tasks against 621 for Gemma4:e2B, and the experiments do not isolate the cause of this difference.
On PRISM-160, the Banking identifiers that P2Skill on Llama3.2:3B transmits in the three runs are all last four card digits, and paraphrases carry these identifiers to the cloud.

\paragraph{Route-only variant (exploratory).}
The route-only (RO) variant P2Skill-RO answers a prompt on the local SLM if P2Skill sets $p_i{=}1$ for any of its sub-tasks or fails to decompose it and otherwise sends the original prompt to the cloud LLM.
P2Skill-RO was defined after the first test results of the frozen P2Skill, which remains the pre-specified method, and its scores are composed from the Local Only and Cloud Only answers and the decomposition flags of the frozen P2Skill runs.
P2Skill-RO is evaluated on the test sets only for an SLM on which the point estimates of its Q and IQ on the validation split exceed those of a random mixture of Local Only and Cloud Only answers at the same leakage.
On the validation split, P2Skill-RO differs from this mixture by 0.021 in Q and 0.063 in IQ on Gemma4:e2B, by 0.033 and 0.151 on Llama3.2:3B, and by 0.005 and $-0.038$ on Qwen3.5:2B, where it fails the condition and is not evaluated on the test sets, and all six intervals include zero.
Its test results are not an independent confirmation, because a rough test-set composition motivated the variant and its full composition for Gemma4:e2B and Llama3.2:3B was computed before the condition was checked.
On Llama3.2:3B, its advantage relative to the mixture comes from sending more prompts, which contain fewer identifiers, to the cloud LLM, and it vanishes at the same share of cloud-routed prompts ($-0.035$ [$-0.069$, 0.002] in validation Q and $-0.002$ [$-0.020$, 0.016] in PUPA Q).
On PUPA, the IQ of P2Skill-RO differs from that of the matched mixture by +0.043 [$-0.014$, 0.101] and +0.214 [0.124, 0.306] on Gemma4:e2B and Llama3.2:3B and from that of Local Only by +0.045 [$-0.018$, 0.110] and +0.467 [0.360, 0.577].

\paragraph{Second judge.}
A second judge, Claude Sonnet (\texttt{claude-sonnet-5-5}), added to the evaluation plan after the GPT-4o judgments had started and before any Claude judgment existed, serves only as a descriptive robustness check, and it is not independent of P2Skill and may favor its outputs, because Claude Sonnet guides the refinement of P2Skill.
The refinement calls Claude Sonnet by a model alias that resolved to this snapshot when the judge was added to the evaluation plan, and the snapshot used during the refinement is not recorded.
The Claude judge applies the same Q and IQ judge prompts and parsers to a fixed random subset of 200 PUPA test queries and to all PRISM-160 prompts, for Local Only on each SLM, Presidio, Presidio+GLiNER, P2Skill on Gemma4:e2B and Llama3.2:3B, PAPILLON on Llama3.2:3B, and, in IQ only, Cloud Only.
The planned Claude judgments of P2Skill on Qwen3.5:2B and of the re-optimized PAPILLON are missing at the time of this analysis, and PRISM, the LDP baselines, and the exploratory variants are not judged.
The Claude Code command-line interface that calls this judge adds a short neutral context, controls neither the temperature nor the output length nor the JSON format, and runs at a high reasoning effort that the command in the evaluation plan did not specify.
The safety filters of Claude refuse 196 of 5,525 PUPA and 100 of 4,402 PRISM-160 judge calls, a non-random exclusion of at most 2 prompts per judged method and benchmark.
The Claude judge is stricter than GPT-4o (Cloud Only IQ 5.95 on the PUPA subset and 6.30 on PRISM-160, against 8.60 and 9.06) and, unlike GPT-4o, rates Local Only on Gemma4:e2B below Cloud Only, which may reflect a ceiling of the GPT-4o judge.
Per-prompt agreement between the two judges ranges from near chance to moderate (Cohen's kappa 0.04 to 0.43 for Q, Spearman correlation 0.33 to 0.80 for IQ).
For P2Skill on Gemma4:e2B, both judges give Q and IQ differences of the same sign and interval class (above, below, or including zero) to Local Only, PAPILLON on Llama3.2:3B, and Presidio+GLiNER on both benchmarks, and both judges rate Local Only on this SLM above P2Skill in Q and IQ.
On Llama3.2:3B, the Q difference between P2Skill and Local Only changes sign, from $-0.112$ [$-0.198$, $-0.025$] by GPT-4o to $+0.102$ [0.041, 0.162] by Claude on the PUPA subset and from $-0.006$ [$-0.108$, 0.095] to $+0.209$ [0.108, 0.310] on PRISM-160.

\subsection{Test Details and Descriptive Results}
\label{app:details}

Brackets give 95\% paired bootstrap intervals of differences, dominance follows the rule of Section~\ref{sec:setup}, and every comparison of this section apart from H1 to H3 is descriptive.
Per prompt, PQ multiplies the IQ of the response by the share of the annotated identifiers of the prompt that the first run does not transmit, which makes it equal to the IQ when no identifier reaches the cloud and zero when every identifier does.
Table~\ref{tab:main} gives the mean PQ of the prompts that contain annotated identifiers, and the product counts a loss of one IQ point the same as the transmission of a tenth of the identifiers of a prompt at an IQ of 10, a weighting that the experiments do not validate.
This paper omits several analyses that the evaluation plan lists for reporting, namely the cost, coverage, PII-free fidelity, and deduplicated analyses, the leakage variants, and the four-way split of the Q judgments.

\paragraph{Confirmatory tests.}
H1 counts 213 queries on which only P2Skill preserves quality against 70 on which only Presidio+GLiNER does (Holm-adjusted $p = 1.2 \times 10^{-17}$), H2 counts 310 against 41 such queries ($p = 9.0 \times 10^{-52}$), and H3 counts 36 queries on which only P2Skill preserves quality against 148 on which only Local Only does ($p = 2.5 \times 10^{-17}$).
H1 and H2 pass and H3 fails.
A rule of the evaluation plan that was fixed before the final evaluation then applies the outcome for a failed H3, which allows no utility claim relative to Local Only, instead of the outcome for passing H1 and H2.

\paragraph{Descriptive comparisons on PUPA.}
On PUPA, P2Skill preserves quality more often than Presidio, Presidio+GLiNER, PRISM, and both LDP baselines on each SLM, by intervals that exclude zero, and dominates each of them on each SLM except Presidio+GLiNER on Qwen3.5:2B and Llama3.2:3B, but its dominance of PRISM on these two SLMs rests on Q alone.
On Gemma4:e2B, P2Skill transmits 7.1 [2.5, 11.7] percentage points fewer annotated identifiers than Presidio+GLiNER.
Presidio and Presidio+GLiNER on Qwen3.5:2B and Llama3.2:3B and Selective LDP on Gemma4:e2B reach a higher IQ than P2Skill, and Presidio+GLiNER transmits fewer identifiers on Llama3.2:3B (5.5 [0.6, 10.7] percentage points).
The IQ of P2Skill stays 1.11 to 1.56 points below that of Cloud Only, and its first-run J-Leak differs from its first-run Leak by at most 1.8 percentage points on each SLM.
The IQ of Local Only is 0.02, 0.69, and 1.33 points below that of Cloud Only on Gemma4:e2B, Qwen3.5:2B, and Llama3.2:3B, and the IQ of P2Skill is below that of Local Only on each SLM, by intervals that exclude zero on Gemma4:e2B and Qwen3.5:2B and include zero on Llama3.2:3B.
Against PAPILLON on Llama3.2:3B, P2Skill has a higher Q and IQ at a leakage difference that includes zero, but its higher leakage point estimate keeps it from dominating PAPILLON.
Against the re-optimized PAPILLON and the exploratory released-prompt PAPILLON on Gemma4:e2B and Qwen3.5:2B, P2Skill has a lower leakage point estimate and a lower IQ, its Q is lower on Gemma4:e2B and differs by an interval that includes zero on Qwen3.5:2B, and neither method dominates the other on the same SLM.

\begin{table*}[tp]
\centering
\scriptsize
\setlength{\tabcolsep}{4pt}
\begin{tabular}{@{}l ccc ccc ccc@{}}
\toprule
& \multicolumn{3}{c}{\textbf{Gemma4:e2B}} & \multicolumn{3}{c}{\textbf{Qwen3.5:2B}} & \multicolumn{3}{c}{\textbf{Llama3.2:3B}} \\
\cmidrule(lr){2-4} \cmidrule(lr){5-7} \cmidrule(lr){8-10}
\textbf{Method} & Med. & Bank. & Both & Med. & Bank. & Both & Med. & Bank. & Both \\
\midrule
Cloud Only & 0.00 & 0.00 & 0.00 & 0.00 & 0.00 & 0.00 & 0.00 & 0.00 & 0.00 \\
Presidio & 7.47 & 2.58 & 5.03 & 7.47 & 2.58 & 5.03 & 7.47 & 2.58 & 5.03 \\
Presidio+GLiNER & 7.88 & 8.50 & 8.19 & 7.88 & 8.50 & 8.19 & 7.88 & 8.50 & 8.19 \\
\midrule
Local Only & 8.62 & 9.53 & 9.07 & 8.28 & 8.95 & 8.61 & 7.10 & 7.85 & 7.47 \\
Uniform LDP & 3.15 & 2.48 & 2.81 & 2.67 & 2.23 & 2.45 & 2.60 & 2.31 & 2.46 \\
Selective LDP & 8.90 & 0.00 & 4.45 & 7.85 & 0.00 & 3.92 & 7.97 & 0.00 & 3.99 \\
PRISM & 0.00 & 3.77 & 1.89 & 0.00 & 4.95 & 2.48 & 0.00 & 4.28 & 2.14 \\
PAPILLON & 1.77 & 0.34 & 1.06 & 0.23 & 1.34 & 0.78 & 7.45 & 7.80 & 7.62 \\
PAPILLON (released prompt)$^{\dagger}$ & 8.95 & 9.03 & 8.99 & 8.15 & 6.44 & 7.29 & -- & -- & -- \\
P2Skill & 8.75 & 9.47 & 9.11 & 7.33 & 6.78 & 7.05 & 0.65 & 3.48 & 2.06 \\
P2Skill-RO$^{\dagger}$ & 8.62 & 9.53 & 9.07 & -- & -- & -- & 0.95 & 7.85 & 4.40 \\
\bottomrule
\end{tabular}
\caption{Privacy-quality score (PQ) on the PII-containing PRISM-160 prompts, where Med.\@ and Bank.\@ denote the Medical and Banking prompts.}
\label{tab:pq}
\end{table*}

\begin{table*}[tp]
\centering
\scriptsize
\setlength{\tabcolsep}{4pt}
\begin{tabular}{@{}ll cc cccc cccc@{}}
\toprule
& & \multicolumn{2}{c}{\textbf{Leak (count)}$\downarrow$} & \multicolumn{4}{c}{\textbf{Q}$\uparrow$} & \multicolumn{4}{c}{\textbf{IQ}$\uparrow$} \\
\cmidrule(lr){3-4} \cmidrule(lr){5-8} \cmidrule(lr){9-12}
\textbf{Local SLM} & \textbf{Method} & Med./40 & Bank./47 & Med. & Bank. & Tour. & Gen. & Med. & Bank. & Tour. & Gen. \\
\midrule
\multirow{3}{*}{--} & Cloud Only & 40 & 47 & -- & -- & -- & -- & 8.93 & 9.03 & 8.90 & 9.40 \\
 & Presidio & 1 & 32 & 0.275 & 0.150 & 0.150 & 0.800 & 7.67 & 8.95 & 6.60 & 8.95 \\
 & Presidio+GLiNER & 0 & 2 & 0.350 & 0.075 & 0.250 & 0.625 & 7.88 & 8.95 & 7.15 & 8.05 \\
\midrule
\multirow{7}{*}{Gemma4:e2B} & Local Only & 0 & 0 & 0.850 & 1.000 & 0.925 & 0.975 & 8.62 & 9.53 & 8.93 & 9.38 \\
 & Uniform LDP & 27.3 & 33 & 0.800 & 0.975 & 0.400 & 0.750 & 8.72 & 8.93 & 7.30 & 8.20 \\
 & Selective LDP & 0 & 47 & 0.600 & 0.850 & 0.625 & 0.825 & 8.90 & 8.85 & 8.25 & 8.43 \\
 & PRISM & 40 & 15 & 0.075 & 0.125 & 0.075 & 0.250 & 6.60 & 5.38 & 5.90 & 5.65 \\
 & PAPILLON & 30.3 & 45.7 & 1.000 & 0.875 & 0.975 & 0.950 & 8.97 & 9.00 & 9.00 & 9.22 \\
 & PAPILLON (released prompt)$^{\dagger}$ & 0 & 0 & 0.975 & 1.000 & 0.950 & 0.900 & 8.95 & 9.03 & 8.97 & 9.25 \\
 & P2Skill & 0 & 0 & 0.575 & 0.850 & 0.575 & 0.900 & 8.75 & 9.47 & 7.75 & 9.30 \\
\midrule
\multirow{7}{*}{Qwen3.5:2B} & Local Only & 0 & 0 & 0.900 & 1.000 & 0.550 & 0.750 & 8.28 & 8.95 & 5.62 & 8.57 \\
 & Uniform LDP & 27.3 & 33 & 0.375 & 0.175 & 0.425 & 0.525 & 7.55 & 8.25 & 6.30 & 7.35 \\
 & Selective LDP & 0 & 47 & 0.250 & 0.050 & 0.375 & 0.550 & 7.85 & 8.62 & 6.55 & 8.20 \\
 & PRISM & 40 & 15 & 0.775 & 0.400 & 0.075 & 0.325 & 7.12 & 6.72 & 4.15 & 5.42 \\
 & PAPILLON & 39 & 38 & 1.000 & 0.800 & 0.900 & 0.800 & 8.93 & 8.70 & 8.22 & 8.47 \\
 & PAPILLON (released prompt)$^{\dagger}$ & 1 & 13 & 0.875 & 0.500 & 0.925 & 0.675 & 8.32 & 8.75 & 8.55 & 7.67 \\
 & P2Skill & 0.3 & 8.3 & 0.550 & 0.900 & 0.475 & 0.750 & 7.33 & 8.78 & 6.15 & 8.68 \\
\midrule
\multirow{6}{*}{Llama3.2:3B} & Local Only & 0 & 0 & 0.400 & 0.800 & 0.925 & 0.750 & 7.10 & 7.85 & 8.50 & 8.03 \\
 & Uniform LDP & 27.3 & 33 & 0.275 & 0.125 & 0.625 & 0.475 & 7.60 & 8.57 & 8.28 & 7.45 \\
 & Selective LDP & 0 & 47 & 0.250 & 0.050 & 0.675 & 0.650 & 7.97 & 8.85 & 8.55 & 8.30 \\
 & PRISM & 40 & 15 & 0.175 & 0.075 & 0.250 & 0.250 & 7.12 & 6.35 & 5.70 & 5.97 \\
 & PAPILLON & 0 & 2 & 0.225 & 0.350 & 0.375 & 0.350 & 7.45 & 8.25 & 6.85 & 6.15 \\
 & P2Skill & 33.3 & 24.3 & 0.825 & 0.300 & 0.925 & 0.800 & 8.43 & 8.95 & 8.62 & 8.28 \\
\bottomrule
\end{tabular}
\caption{Per-domain results on PRISM-160 for the Medical (Med.), Banking (Bank.), Tourism (Tour.), and General knowledge (Gen.)\ domains.}
\label{tab:per-domain-main}
\end{table*}

\begin{table*}[tp]
\centering
\scriptsize
\setlength{\tabcolsep}{4pt}
\begin{tabular}{@{}ll ccccc ccccc@{}}
\toprule
& & \multicolumn{5}{c}{\textbf{PRISM-160}} & \multicolumn{5}{c}{\textbf{PUPA test}} \\
\cmidrule(lr){3-7} \cmidrule(lr){8-12}
\textbf{Local SLM} & \textbf{Method} & Leak$\downarrow$ & J-Leak$\downarrow$ & Q$\uparrow$ & IQ$\uparrow$ & PQ$\uparrow$ & Leak$\downarrow$ & J-Leak$\downarrow$ & Q$\uparrow$ & IQ$\uparrow$ & PQ$\uparrow$ \\
\midrule
\multirow{2}{*}{Gemma4:e2B} & PAPILLON (released prompt) & 0.0 & 0.0 & 0.956 & 9.05 & 8.99 & 28.7 & 29.6 & 0.857 & 8.56 & 5.33 \\
                           & P2Skill-RO                 & 0.0 & -- & 0.938 & 9.12 & 9.07 & 11.3 & -- & 0.938 & 8.68 & 7.00 \\
\midrule
Qwen3.5:2B                 & PAPILLON (released prompt) & 16.1 & 16.1 & 0.744 & 8.32 & 7.29 & 31.4 & 33.5 & 0.657 & 7.41 & 4.46 \\
\midrule
Llama3.2:3B                & P2Skill-RO                 & 38.3 & -- & 0.912 & 8.56 & 4.40 & 19.7 & -- & 0.770 & 7.80 & 4.90 \\
\bottomrule
\end{tabular}
\caption{Leakage and quality of the exploratory variants, where P2Skill-RO denotes the route-only variant of P2Skill.}
\label{tab:exploratory}
\end{table*}

\paragraph{Per-domain results on PRISM-160.}
Table~\ref{tab:pq} splits the PRISM-160 PQ of Table~\ref{tab:main} into the 40 Medical and the 40 Banking prompts and adds Local Only, whose PQ equals its IQ on these prompts because it transmits nothing.
On Gemma4:e2B, the PQ differences between P2Skill and Local Only and between P2Skill and the exploratory released-prompt PAPILLON, 0.038 [$-0.150$, 0.250] and 0.125 [$-0.025$, 0.263], include zero.
Table~\ref{tab:per-domain-main} gives the leakage, Q, and IQ of each PRISM-160 domain, where Leak counts the transmitted annotated identifiers as the mean of the runs, as in Table~\ref{tab:main}.
In both tables, Cloud Only, Presidio, and Presidio+GLiNER use no SLM, a dash marks a value that does not apply or a variant that is not run on the SLM, and a dagger marks an exploratory variant.

\paragraph{Exploratory variants.}
Table~\ref{tab:exploratory} gives the results of the two exploratory variants in the columns of Table~\ref{tab:main}, namely PAPILLON using the released prompt on Gemma4:e2B and Qwen3.5:2B and the route-only variant P2Skill-RO on Gemma4:e2B and Llama3.2:3B.
A dash in Table~\ref{tab:exploratory} marks a value that is not computed.
PAPILLON using the released prompt reaches a PQ that differs from the PQ of P2Skill by paired intervals that include zero on both benchmarks and dominates P2Skill on Gemma4:e2B on PRISM-160.
P2Skill-RO reaches a higher PUPA PQ than P2Skill, 7.00 against 5.61 on Gemma4:e2B and 4.90 against 3.90 on Llama3.2:3B.
The Q of P2Skill-RO is not comparable to the Q of the other methods, because a prompt that it sends to the cloud receives the Cloud Only response, which is the target response of Q.

\subsection{Cross-SLM Skill Transfer}
\label{app:transfer}

In a pre-specified descriptive analysis, the frozen skills of each local SLM run on the other two SLMs, the cloud-bound requests do not reach the cloud LLM, quality is not measured, and each cell of Table~\ref{tab:transfer} gives the mean of three runs, in italics for an SLM on its own skills.
After timeouts of the Ollama server, one PUPA test run of Llama3.2:3B lacks one prompt on the skills of Gemma4:e2B and another lacks three prompts on the skills of Qwen3.5:2B, these prompts are not retried, and the leakage of these two runs covers the remaining prompts.
No transfer meets the applicability criterion of the evaluation plan, which requires a PUPA test leakage within the run range of the executing SLM on its own skills (15.2--15.9\%, 25.6--29.4\%, and 27.5--29.9\% for Gemma4:e2B, Qwen3.5:2B, and Llama3.2:3B) and the coverage and retention bounds of the skill selection in every run.
The three transfers in which Gemma4:e2B runs the skills of Qwen3.5:2B and Llama3.2:3B runs the skills of either other SLM leak less than this range but miss at least one bound, as do Gemma4:e2B and Qwen3.5:2B on their own skills, and the other three transfers leak more than this range, including the two marked by an asterisk that meet every bound in every run.

\begin{table}[t]
\centering
\footnotesize
\setlength{\tabcolsep}{4pt}
\begin{tabular}{@{}lccccccc@{}}
\toprule
& \multicolumn{3}{c}{\textbf{PUPA test}} & & \multicolumn{3}{c}{\textbf{PRISM-160}} \\
\cmidrule{2-4} \cmidrule{6-8}
\textbf{Skills of} & G4 & Q3.5 & L3.2 & & G4 & Q3.5 & L3.2 \\
\midrule
Gemma4:e2B  & \textit{15.6} & 53.6 & 27.1 & & \textit{0.0} & 30.7 & 16.1 \\
Qwen3.5:2B  & 11.4 & \textit{27.3} & 23.4 & & 0.0 & \textit{10.0} & 36.0 \\
Llama3.2:3B & 62.5$^{\ast}$ & 67.0$^{\ast}$ & \textit{28.3} & & 4.2$^{\ast}$ & 96.9$^{\ast}$ & \textit{66.3} \\
\bottomrule
\end{tabular}
\caption{String-match leakage (\%) of cross-SLM skill transfer, where rows give the skills, columns give the executing SLM, and G4, Q3.5, and L3.2 denote Gemma4:e2B, Qwen3.5:2B, and Llama3.2:3B.}
\label{tab:transfer}
\end{table}

\end{document}